\documentclass{sigkddExp}\usepackage[]{graphicx}\usepackage[]{color}
\makeatletter
\def\maxwidth{ %
  \ifdim\Gin@nat@width>\linewidth
    \linewidth
  \else
    \Gin@nat@width
  \fi
}
\makeatother

\definecolor{fgcolor}{rgb}{0.345, 0.345, 0.345}

\usepackage{framed}
\makeatletter
 {\par\unskip\endMakeFramed%
 \at@end@of@kframe}
\makeatother

\definecolor{shadecolor}{rgb}{.97, .97, .97}
\definecolor{messagecolor}{rgb}{0, 0, 0}
\definecolor{warningcolor}{rgb}{1, 0, 1}
\definecolor{errorcolor}{rgb}{1, 0, 0}

\usepackage{alltt}
\usepackage[utf8]{inputenc} 
\usepackage[T1]{fontenc}    
\usepackage{hyperref}       
\usepackage{url}            
\usepackage{booktabs}       
\usepackage{amsfonts}       
\usepackage{nicefrac}       
\usepackage{microtype}      
\usepackage{float}
\usepackage{amsmath}
\usepackage{scrextend}
\usepackage{lipsum}
\usepackage{algorithm2e}
\usepackage{caption}
\usepackage{booktabs}
\usepackage{multirow}
\usepackage{siunitx}
\usepackage{subfigure}
\usepackage{eurosym}
\usepackage{bm}
\usepackage{graphicx}
\RestyleAlgo{boxed}

\usepackage{tikz}
\usetikzlibrary{fit,positioning}
\tikzset{
  varnode/.style={draw, minimum width=10mm, shape=circle, thick, black},
  dagconn/.style={arrows=->, black, thick},
  plate/.style={draw, shape=rectangle, rounded corners=0.5ex, thick,
    minimum width=2.1cm, text width=2.1cm, align=right, inner sep=10pt, inner ysep=10pt,label={[xshift=-14pt,yshift=14pt]south east:#1}}
}
\IfFileExists{upquote.sty}{\usepackage{upquote}}{}

\begin{document}
%

\title{Scalable Decisions using a Bayesian Decision-Theoretic Approach}

\numberofauthors{2}

\author{
%
\alignauthor Hoiyi Ng \\
       \affaddr{Amazon.com, Inc.}\\
       \affaddr{Seattle, WA}\\
\alignauthor Guido Imbens\titlenote{Guido Imbens holds concurrent appointments as an Amazon Scholar and as a Professor of Economics at Stanford University. This paper describes work performed at Amazon.}\\
       \affaddr{Stanford University}\\
       \affaddr{Palo Alto, CA}\\
}
\maketitle

\begin{abstract}
Randomized controlled experiments are commonly employed
to assess impacts of new policies (e.g., a new pricing strategy,
and a new inventory placement mechanism, etc.) on a set
of performance metrics (e.g., revenue, profit, and customer
experience, etc.). The goal of running such experiments is to
decide whether to launch a new policy. Often, the impacts
of a policy on different metrics are assessed independently
despite the correlation between metrics. In addition, the
estimated impacts are not universally beneficial. For example,
a new policy can have a positive impact on a revenue
metric, and a negative impact on a customer
experience metric. As a result, decision making requires
human intervention and judgement, which hinders scalability and policy development.
We propose a Bayesian decision-theoretic approach for
decision-making under multiple objectives (performance metrics
defined by experimenters) and corresponding trade-offs
(loss function defined by experimenters), where we compare
expected risks under different decisions. We formalize a decision
making framework based on Bayesian decision theory,
allowing experimenters to combine a desired loss function
with observed evidence in decision making. We use hierarchical
models to obtain prior information for policy impacts
(treatment effects) based on learnings from historical experiments.
We show through real and simulated Amazon supply
chain experiments that compared to traditional null hypothesis
statistical testing (NHST), our approach: 1) increases
estimation efficiency through leveraging an informative hierarchical prior,
and 2) greatly simplifies the decision making process for
experimenters and stakeholders while systematically incorporating
Amazon’s business preferences and costs for more
comprehensive decisions.
\end{abstract}

\section{Introduction}
\vspace{5mm}
In-production experiments are essential tools for understanding
the downstream impacts of innovations in Amazon’s
supply chain. The goal of such experiments is often to decide
whether to roll out a new policy. In many settings (e.g.,
testing layouts of websites, advertising, etc), such
decisions are made to minimize regret in an optimization
problem. For example, more webpage customers may be assigned
to the ‘best policy’ through different sampling mechanisms
such as Thompson sampling in multi-armed bandit experiments \cite{thompson1933likelihood}.
Further examples of formulating decision making as
multi-armed bandit experiments are presented in
\cite{hill2017efficient,scott2015multi}. 
However,
to make optimal decisions using multi-armed bandits,
a fast feedback loop is often required. Unlike website related
policies such as changing the font or color of texts, which
immediately affects downstream metric such as conversion rate,
supply chain policies often require a much longer lead time
before their impacts propagate to downstream metrics. For
example, if a new policy changes inventory buying behavior,
it will take weeks before these changes
can be reflected in revenue, after accounting for the time
required to buy the inventory using the new policy, the time
required for the vendors to send the inventory, and the time
required for customers to purchase the items.\\

Therefore, decision making (launch or roll-back a policy
being assessed) for Amazon’s supply chain experiments has relied
traditionally on NHST where the
impacts of a policy on different metrics are independently
assessed and decisions are based on statistical significance.
Due to many well documented issues with p-values
\cite{blume2003your, sterne2001sifting, wasserstein2016asa},
we moved away from p-values towards an emphasis
on confidence intervals and more recently a Bayesian probabilistic
approach to aid decision making. However, since
we currently have no methodology for combining treatment
effect estimates across multiple metrics, experimenters (e.g.,
policy developers and stakeholders) still often struggle when
deciding whether to launch a new policy based on statistical
results. Challenges include: 1) inaccurate
and/or inconsistent interpretation of statistical results and
associated uncertainty, 2) lack of a universal business performance
measure, 3) policies not being ubiquitously good, and
4) inconsistent business objectives across policy owners and
stakeholders. Due to the physics of the supply chain, policy
impacts are often not universally beneficial (e.g, a policy that
increases revenue might also require more investment), creating
conflict between stakeholders and experimenters when
making decisions. As such, it is necessary to incorporate
trade-offs (loss function) between these performance metrics
in decisions.\\

To scale and streamline the decision making process, we
establish a decision making framework based on Bayesian
decision theory to incorporate multiple objectives, their corresponding
trade-offs, and various related costs. Bayesian
decision theory attempts to combine statistical information
from a sample (e.g., efficacy of a drug, impact of a new
advertising campaign) with other relevant aspects of the
problem to derive the best decision. Often, the consequences
of different decisions, which can be quantified by the loss
incurred, are considered. Aside from the loss function, prior
information on the parameter of interest is also considered \cite{berger2013statistical}.
Various approaches for obtaining such prior information are
discussed in \cite{gelman2013bayesian}, with hierarchical models being one of the
most commonly used approaches. Hierarchical priors are
often used in sharing statistical strengths and feature selection
in machine learning. Benefits of hierarchical
priors, such as robustness and sharing of statistical strength,
are discussed in \cite{gelman2013bayesian, huggins2015risk}. Further examples where hierarchical
modeling is critical to obtaining high-quality inferences are
provided in \cite{gelman2006data, gelman2013bayesian}. By combining prior information, statistical
information observed from an experiment, and a loss
function using Bayesian decision theory, our approach allows
experimenters to: 1) define a flexible loss function that
incorporates trade-offs between different objectives (business
preferences) and costs, 2) leverage learnings from similar
historical experiments, and 3) make decisions based on automated
recommendations. \\

This paper is organized as follows: we formulate our decision
making framework in Section 2. In Section 3, we describe
how to construct a prior distribution for treatment eﬀects
on a given set of performance metrics using hierarchical
models. In Section 4, we quantify the advantages (in terms
of estimation eﬃciency) of our proposed framework through
numerical studies. In Section 5, we illustrate how to use
this framework in practice by showing expected risks under
diﬀerent decisions and corresponding decision spaces. In
Section 6, we discuss diﬀerent approaches to defining a loss
function. We devote Section 7 to discussion and future work.

\section{Decision-Making Framework}
\vspace{5mm}
\subsection{Background}
\vspace{5mm}
Bayesian decision theory is a fundamental statistical approach to quantifying consequences (risks)
between different decisions (i.e., launch vs. roll-back) using probability and the costs that
accompany such decisions \cite{berger2013statistical}.
It is often used in decision making under uncertainty. Making decisions
using Bayesian decision theory requires three main components: 1) a decision space
$\mathcal{A}$ = $\{$launch, roll-back$\}$, 2) an information space offering observed information,
i.e, impacts of a new policy \textbf{w},
and 3) a loss function $l(a|\textbf{w})$ that quantifies the costs of each action given the
impacts of a policy. This loss function ties together the decision space and information space.
If the true treatment effects (\textbf{w}) are known, we can simply choose a decision that gives a
smaller loss, $\text{argmin}_{a}l(a|\textbf{w})$. However, we almost never observe the true treatment
effects. Instead, based on observed data \textbf{x} collected in experiments, we can construct a
posterior distribution of the treatment effects $P(\textbf{w}|\textbf{x})$. To incorporate uncertainty, 
we make decisions by comparing expected risks under different decisions:

\begin{equation}
\mathcal{R}(a|\textbf{x}) = \int \ell(a,\mathbf{w})P(\mathbf{w}|\mathbf{x})dw.
\end{equation}

\subsection{Formulation}
\vspace{5mm}
In our decision making framework, we make launch decisions by comparing expected risks
of launch versus roll-back actions. We consider an action space
$\mathcal{A} = \{a_0, a_1\}$, where $a_0$ and $a_1$
represent roll-back and launch; an information space, $\textbf{w}\in \mathbb{R}^n$,
of unobserved treatment
effects on the success metrics, where $n$ denotes the number of metrics chosen by the experimenter
to define success (i.e., success metrics); corresponding observed information
$\mathbf{x} \in \mathbb{R}^n$
collected from experiments (i.e., estimated treatment effects);
and a loss function jointly defined by costs (estimated roll-back cost $c_0$ and launch cost $c_1$)
and trade-offs $\Lambda \in \mathbb{R}^n$, a vector of trade-offs between
treatment effects reflecting the business' preference. For example, let $w_1$, $w_2$, and $w_3$ denote
the unknown impact of a policy on revenue, profit, and investment, respectively,
and let $\Lambda = \{10, 1, -10000\}$. This trade-off vector suggests
that a unit increase profit is worth 10
units increase in revenue and \$10000 investment. In other words, each element of $\Lambda$
represents the relative value of the treatment effect on the
corresponding metric. Our framework can flexibly incorporate
any loss functions. The trade-off vector stated above
implicitly assumes a linear loss function.\\

We define the expected risk of an action given evidence as the average loss over the posterior
distribution of the treatment effects:
\begin{equation}
\mathcal{R}(a_i | \mathbf{x}) = \int \ell(a_i,\mathbf{w})P(\mathbf{w}|\mathbf{x})dw
\end{equation}

\begin{equation}
P(\mathbf{w}|\mathbf{x}) \propto P(\mathbf{x}|\mathbf{w})\cdot P(\mathbf{w})
\end{equation}
where $P(\mathbf{w})$ and $P(\mathbf{x}|\mathbf{w})$ are the multivariate prior for the treatment
effects and multivariate likelihood, respectively. $\ell(a_i,\mathbf{w})$ is the loss function
defined by stakeholders. In
this paper, we use a simple linear loss function to illustrate.\\

Let $\Lambda \in \mathbb{R}^n$ be the trade-off vector.
This implies that the linear loss function $\ell(a_i,\mathbf{w})$ (the loss incurred by
launching (roll-back) the policy given the unobserved treatment effects) is as follows:

\begin{equation}
\ell(a_i,\mathbf{w}) = 
  \begin{cases}
  \sum_j\lambda_jw_j + c_0 & i = 0 \\
  -\sum_j\lambda_jw_j + c_1 & i = 1
  \end{cases}
\end{equation}
where $w_i$'s are elements of \textbf{w}. In other words, the expected loss is a linear
combination of the treatment effects on the success metrics. If $\lambda_i = 1$ for all $i$, the
expected loss is a simple average of the treatment effects.
On the other hand, if $\lambda_1 = 1$ and $\lambda_i = 0$ for $i\neq 1$, the expected loss is only driven by the
treatment effect on a single metric.\\

Therefore, the expected risk for a launch action accounting for costs is:
\begin{equation}
\mathcal{R}(a_1 | \mathbf{x}) = -\int \sum_j\lambda_jw_jP(\mathbf{w}|\mathbf{x})d\mathbf{w} + c_1
\label{eq:riska1}
\end{equation}

Similarly, the expected risk associated with a roll back action accounting for costs is:
\begin{equation}
\mathcal{R}(a_0 | \mathbf{x}) = \int \sum_j\lambda_jw_jP(\mathbf{w}|\mathbf{x})d\mathbf{w} + c_0
\label{eq:riska0}
\end{equation}

A policy should be launched if the expected risk associated with launching is lower than that 
associated with rolling back ($\mathcal{R}(a_1 | \mathbf{x}) < \mathcal{R}(a_0 | \mathbf{x})$). \\

In addition to enabling the consideration of trade-offs and costs. These trade-offs
naturally allows experimenters to apply `guardrails’ on certain metrics. For example, if one does
not want to launch the policy if a given metric is negatively impacted, one can simply inflate the
value of this given metric relative to other metrics. The posterior
distribution of the treatment effects allows experimenters to understand the joint probability of
success. In other words, it answers the question ``what is the probability that the impacts on both
revenue and customer experience will both be greater than 0?”.

\subsection{Calculating Expected Risks}
\vspace{5mm}
To calculate the expected risks, we can sample from the posterior distribution of the treatment
effects, compute the risk of each sample, then average over the risks from all samples. However,
when the loss function is linear, we can use the following trick to further reduce
computational time. Let 
$\bm{\Lambda} = [\lambda_1, \ldots, \lambda_i]$ be the trade-off vector corresponding to metrics 
$1, \cdots, i$ on which the unobserved treatment effects follow a multivariate normal distribution
with mean $\bm{\tau}$ and variance $\bm{\Delta}$ ($P(\textbf{w}|\textbf{x}) \sim \mathcal{N}(\bm{\tau}, \bm{\Delta}))$. The 
expected risks defined in equation (\ref{eq:riska1}) and (\ref{eq:riska0}) can be as considered
affine transformations of the unobserved treatment effects $\textbf{w}$, yielding

\begin{align}
\mathcal{R}(a_1|\textbf{x}) &= -\mathcal{G}(\bm{\Lambda})^T\bm{\tau} + c_1,
\end{align}

and, 
\begin{align}
\mathcal{R}(a_0|\textbf{x}) = \mathcal{G}(\bm{\Lambda})^T\bm{\tau} + c_0,
\end{align}

where $\mathcal{G}(x)$ is a transformation on a vector $x \in \mathbb{R}^{n}$ such that the
$i^{th}$ element $x_i$ of $x$ is given by:

\begin{equation}
\frac{1/x_i}{\sum_{j = 1}^{n} 1/|x_j|}.
\end{equation}

For example, if $\Lambda = [1, 99]$, then $\mathcal{G}(\Lambda) = [0.99, 0.01]$.
Such transformation allows stakeholders to specify intuitive trade-offs (e.g., how much am I 
willing to invest in inventory to gain \$1 in revenue?) while keeping the expected impact portion 
of the risk comparable to the expected costs.

\section{Prior Construction}
\begin{figure}[H]
\centering
\begin{tikzpicture}[scale=0.7, transform shape]
  \node[varnode] (mu) {$\bm{\mu}$, $\bm{\Gamma}$};
  \node[varnode] (tau2) [below=of mu] {$\textbf{w}_2$ };
  \node[varnode] (tau1) [left=of tau2] {$\textbf{w}_1$ };
  \node[varnode, color = white, text = black] (tau3) [right=of tau2] {$\cdots$};
  \node[varnode] (taut) [right=of tau3] {$\textbf{w}_t$ };
  \node[varnode] (x1) [below=of tau1] {$\textbf{x}_1$};
  \node[varnode] (x2) [below=of tau2] {$\textbf{x}_2$};
  \node[varnode, color = white, text = black] (x3) [below=of tau3] {$\cdots$};
  \node[varnode] (xt) [below=of taut] {$\textbf{x}_t$};
  \draw[dagconn] (mu) to (tau1);
  \draw[dagconn] (mu) to (tau2);
  \draw[dagconn] (mu) to (taut);
  \draw[dagconn] (tau1) to (x1);
  \draw[dagconn] (tau2) to (x2);
  \draw[dagconn] (taut) to (xt);
\end{tikzpicture}
\caption{A hierarchical model for treatment effects. $\bm{\mu}$ and $\bm{\Gamma}$ represents the mean and the variance of the treatment effect distribution across similar experiments. $\textbf{w}_i$ represents the
unobserved treatment effects of experiment $i$, and $\textbf{x}_i$ represents the results (estimated
treatment effects) we observe
in experiment $i$. We are interested in the distribution of $\textbf{w}_t$ given information
$x_1, \cdots, x_{t-1}$.}
\label{fig:hierGraph}
\end{figure}
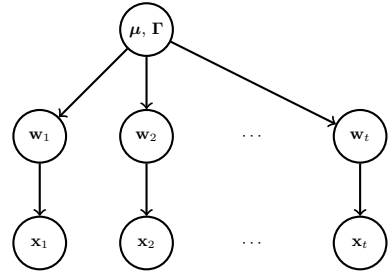

\vspace{5mm}
Traditionally, we discard valuable learning from historical
experiments.
For example, suppose we assessed a policy
with parameter setting $s$ through an experiment 6 months
ago and found that the policy boosted revenue, and now
want to assess the impact of a new policy with parameter
setting $s + 5$. When estimating the impact of this new policy with
parameter setting $s + 5$ in the new experiment, we ignore
our prior knowledge about the impact
of parameter setting $s$, and focus solely on information we
collect during the new experiment. To more efficiently estimate the treatment effects at time $t$
using prior knowledge, we leverage results from experiments
conducted up to time $t - 1$ to construct a prior distribution
for treatment effects of the new policy being assessed by
an experiment at time $t$ through a hierarchical model. A
graphical representation of the hierarchical model is shown in Figure \ref{fig:hierGraph}. We assume
that the unobserved treatment effects $\textbf{w} \in \mathbb{R}^{n\times 1}$ from historical
experiments come from a common distribution with mean $\bm{\mu} \in \mathbb{R}^{n\times 1}$ and
variance $\bm{\Gamma} \in \mathbb{R}^{n\times n}$. Assuming $\bm{\Gamma}$ is known,
let $\textbf{x}_i$ be the estimate of the true treatment effect $\textbf{w}_i$, and define

\begin{align}
\bm{\mu}  & \sim \mathcal{N}(0, \infty) \label{eq:mu_dist}\\
\textbf{w}_i | \bm{\mu} & \sim \mathcal{N}(\bm{\mu}, \mathbf{\Gamma}) \\
\textbf{x}_i | \textbf{w}_i & \sim \mathcal{N}(\textbf{w}_i, \mathbf{\Sigma}_i)
\end{align}

Based on the hierarchical structure, we can derive the prior distribution for the treatment effects
of experiment $t$ as the posterior distribution of the true treatment effects in the experiment at
time $t$ given results from historical experiments up to time $t -1$ ($\textbf{x}_{-t}$):
\begin{align}
P\left(\textbf{w}_t|\textbf{x}_{-t}\right) 
& = \int P(\textbf{w}_t | \bm{\mu}, \textbf{x}_{-t}) \cdot P(\bm{\mu} | \textbf{x}_{-t})d\mu,
\end{align}
where 

\begin{equation}
P(\textbf{w}_t | \bm{\mu}, \textbf{x}_{-t}) = P(\textbf{w}_t | \bm{\mu}) \sim \mathcal{N}(\bm{\mu}, \bm{\Gamma}),
\end{equation}

since $\textbf{w}_t$ and $\textbf{x}_{-t}$ are independent given $\bm{\mu}$, and

\begin{equation}
P(\bm{\mu} | \textbf{x}_{-t}) \sim 
\mathcal{N}\left(\bm{\Lambda}^{-1}\bm{\nu}, \bm{\Lambda}^{-1}\right).
\end{equation}

The mean precision and the precision of the prior distribution of the $t^{th}$ experiment are given
by
\begin{equation}
\bm{\nu} = \sum_{i < t}(\bm{\Sigma}_i + \bm{\Gamma})^{-1}\textbf{x}_i,
\end{equation}

and 

\begin{equation}
\bm{\Lambda} = \sum_{i < t}\left(\bm{\Sigma}_i + \bm{\Gamma} \right)^{-1}, 
\end{equation}

respectively. Thus,

\begin{equation}
P\left(\textbf{w}_t|\textbf{x}_{-t}\right) \sim
\mathcal{N} ((\bm{\Gamma} + \bm{\Lambda}^{-1})^{-1}\bm{\Lambda}^{-1}\nu, (\bm{\Gamma} + \bm{\Lambda}^{-1})^{-1})
\end{equation}

Let $\bm{\Omega} = \bm{\Gamma} + \bm{\Lambda}^{-1}$, the posterior 

\begin{equation}
\label{eq:posterior}
P\left(\textbf{w}_t|\textbf{x}\right) \sim \mathcal{N}(\bm{\tau}, \bm{\Delta})
\end{equation}
 is normally distributed with mean:

\begin{equation}
\bm{\tau} = (\bm{\Sigma}_t^{-1} + \bm{\Omega}^{-1})^{-1}
(\bm{\Sigma}_t^{-1}\textbf{x}_t + \bm{\Omega}^{-1}\bm{\Lambda}^{-1}\bm{\nu}),
\end{equation}

and variance:

\begin{equation}
\bm{\Delta} = (\bm{\Sigma}_t^{-1} + \bm{\Omega}^{-1})^{-1}.
\end{equation}


At this point, we can see that $\bm{\nu}$ can be estimated by
the model while $\bm{\Gamma}$ is treated as known\footnote{A small adjustment can be made to the
model to estimate $\bm{\Gamma}$, which yields a prior following a student's $t$-distribution instead
of a normal distribution.}. We treat $\bm{\Gamma}$ as a tuning parameter that controls the influence
of the prior on the posterior. To determine the value of $\bm{\Gamma}$, we let $\bm{\Gamma} = k\bm{\Theta}$ be the scaled covariance
of the treatment effects (across different experiments) distribution, and estimate the covariance
$\bm{\Theta}$ of the estimated treatment effects $\textbf{x}$ across different experiments
empirically by:

\begin{equation}
\hat{\bm{\Theta}} = \frac{1}{t-1} \bm{X}\bm{X}'
\end{equation}

where $t-1$ is the number of experiments completed up to time $t$, and $\bm{X}$ is a matrix of
estimated treatment effects whose $i^{th}$ column is the estimated treatment effects 
$\textbf{x}_i$ for experiment $i$\footnote{We assume the mean is 0.}. 
It is trivial to see that the smaller the $\bm{\Gamma}$, the more information we borrow from
historical experiments, leading to more shrinkage towards the prior. In
practice, we only care about 3 values of $k$, viz., $k = 0$ (strong
shrinkage), $k = 1$ (moderate shrinkage), and $k = \infty$ (no shrinkage). We show how different
values of $k$ affect the prior and posterior through numerical studies in Section 4. To obtain
the covariance matrix $\bm{\Sigma_i}$ for each experiment $i$, we use bootstrap.

\section{Numerical Studies}
\vspace{5mm}
In this section, we investigate the choice of the tuning
parameter $\bm{\Gamma}$ ($k$) for the prior and the estimation efficiency gain by
leveraging prior information.

\subsection{Priors with Varying Degrees of Shrinkage}
\vspace{5mm}
Figure \ref{fig:chooseK} below shows the estimated prior mean and standard
deviation of 2 metrics (viz., M1 and M2), constructed using all historical experiments,
as a function of $k$. We see that as $k$ increases, the prior mean
is small and relatively stable while the corresponding standard
deviations increase in orders of magnitudes, implying
less shrinkage to the prior mean.

\begin{figure}[H]
\centering
\includegraphics[width=7.5cm, height = 3.5cm]{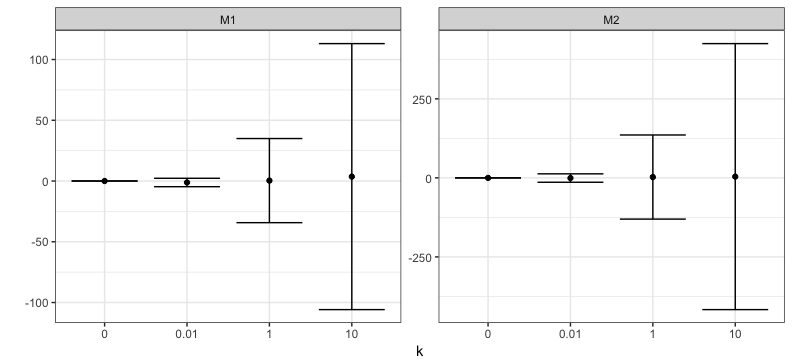}
\caption{Prior mean and standard deviation as a function
of $k$. $k = 1$ produces an unbiased
estimate for the covariance.}
\label{fig:chooseK}
\end{figure}

To investigate how different choices of $k$ affect the prior,
Table 1-2 below show the prior mean and covariance assuming
all historical experiments share one unobserved true
treatment effect ($k = 0 \implies \bm{\Gamma} = 0$). Table 3-4 show the
prior mean and covariance assuming a hierarchical structure
shown in Figure \ref{fig:hierGraph} with $k = 1$. We see that $k = 1$ provides a
reasonable prior while $k = 0$ offers too strong of a prior as all
treatment effects and their corresponding variance (covariance)
are shrunk close to 0. This is unreasonable in practice as
all policies are at least slightly different in nature, and there
are no reasons to believe that all policies will produce the
same impacts.

\vspace{-0.3cm}

\begin{table}[H]
\centering
\caption{Prior mean, $k = 0$}
\begin{tabular}{cc}
\toprule
Metric & Estimate\\
\midrule
M1 & -0.058\\
M2 & -0.024\\
M3 & 0.001\\
\bottomrule
\end{tabular}
\label{tab:meank0}
\end{table}

\vspace{-0.5cm}

\begin{table}[H]
\centering
\caption{Transformed prior covariance, $k = 0$. Diagonal entries are
standard errors, off-diagonal entries
are correlations.}
\begin{tabular}{cccc}
\toprule
   & M1 & M2 & M3\\
\midrule
M1 & 0.01 & 0.06 & 0.09\\
M2 & 0.06 & 0.02 & 0.03\\
M3 & 0.09 & 0.03 & 0.00\\
\bottomrule
\end{tabular}
\label{tab:covk0}
\end{table}

\vspace{-0.5cm}

\begin{table}[H]
\centering
\caption{Prior mean, $k = 1$}
\begin{tabular}{cc}
\toprule
Metric & Estimate\\
\midrule
M1 & 2.6\\
M2 & 131.8\\
M3 & 0.003\\
\bottomrule
\end{tabular}
\label{tab:meank0}
\end{table}

\vspace{-0.5cm}
\begin{table}[H]
\centering
\caption{Transformed prior covariance, $k = 1$. Diagonal entries are
standard errors, off-diagonal entries
are correlations.}
\begin{tabular}{cccc}
\toprule
   & M1 & M2 & M3\\
\midrule
M1 & 133.1 & 0.04 & -0.04\\
M2 & 0.04 & 1278.9 & 0.01\\
M3 & -0.04 & 0.01 & 0.02\\
\bottomrule
\end{tabular}
\label{tab:covk0}
\end{table}

\subsection{Posteriors with Varying Degrees of Shrinkage}
\begin{figure}[H]
\centering
\includegraphics[width=8.5cm, height = 4cm]{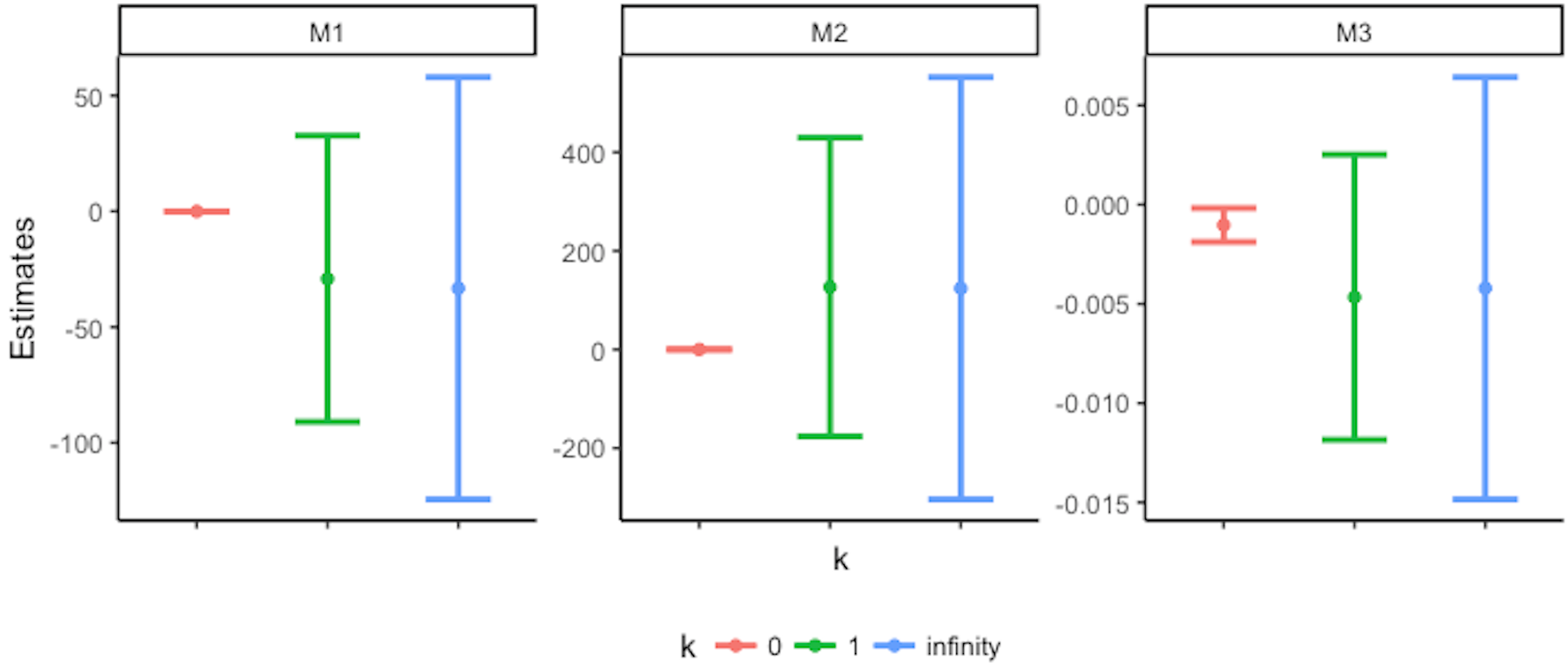}
\caption{Treatment effects estimates and corresponding 95\%
credible intervals under different choices of $k$}
\label{fig:postComp}
\end{figure}

\vspace{5mm}
To investigate how different choices $k$ affect the posterior,
we look at the posterior distribution of a pseudo-synthetic
experiment under different values of $k$ of three metrics (viz.,
M1, M2, and M3). Figure \ref{fig:postComp} shows estimated treatment
effects and corresponding 95\% confidence intervals ignoring
prior information (this is equivalent to setting $k = \infty$), and
estimated treatment effects and corresponding 95\% credible
intervals with a strong prior ($k = 0$), and a moderate
prior with $k = 1$. 

Although the true treatment effects are unknown in these pseudo-synthetic experiments,
we can use treatment effect estimates from a model that ignores prior information
as the benchmark since they are unbiased estimates of the true treatment
effects. We see that using an unbiased estimate for the variance $\bm{\Gamma}$ produces
a reasonable posterior as we observe that the point
estimates are similar for $k = \infty$ and $k = 1$. However, using
prior information produces shorter credible intervals. In
contrast, we see again that assuming all historical
experiments share the same unobserved true treatment
effects is unreasonable and likely introduces large bias.  

\subsection{Estimation Efficiency on Real and Simulated Data}
\begin{figure}[H]
\centering
\includegraphics[width=5.5cm, height=3.5cm]{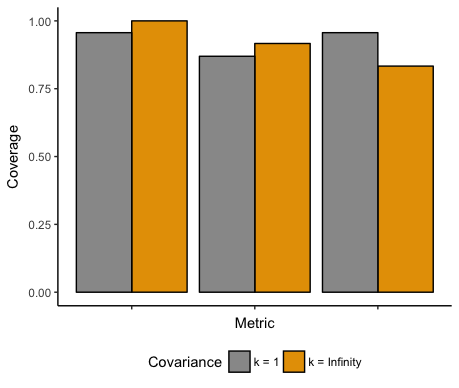}
\caption{Mean coverage across 20 experiments under different choices of $k$.}
\label{fig:coverage}
\end{figure}

\vspace{5mm}
To compare the estimation performance of ignoring prior
information and our proposed Bayesian approach leveraging
learning from historical experiments, we survey 70+ real supply
chain experiments. By incorporating prior information
using our approach, we find that the marginal significance
rate increases by 38\%.\footnote{We define significance as when
the treatment effect estimates are outside of the 95\% credible
intervals.} To further investigate the estimation
efficiency gain from using prior information, we repeat this
exercise on simulated data from 20 supply chain experiments.
In each experiment, we randomly shuffle the treatment assignments
(the true treatment effect is 0 by construction), then estimate
the treatment effects using two different approaches (e.g.,
using prior information by setting $k = 1$ and ignoring prior
information by setting $k = \infty$). Table \ref{tab:mse} and Figure
\ref{fig:coverage} show
the mean square error\footnote{Since the true treatment effects are 0 by 
design, we can calculate MSE by taking the mean of the squared treatment
effect estimates.} and coverage\footnote{Coverage is the proportion of times
that the 95\% credible intervals cover the true treatment effects 0.}
of the two different
approaches on three performance metrics (viz., M1, M2,
and M3), respectively. We observe that leveraging prior
information results in a large efficiency gain while preserving
coverage.

\vspace{-0.2cm}

\begin{table}[H]
\centering
\caption{MSE across 20 sets of simulated data.}
\begin{tabular}{ccc}
\toprule
   & $k = 1$ & $k = \infty$\\
\midrule
M1 & 20 & 16800\\
M2 & 9651 & 690919\\
M3 & $5e^{-5}$ & $7e^{-4}$\\
\bottomrule
\end{tabular}
\label{tab:mse}
\end{table}

\vspace{-0.2cm}
Finally, we investigate the changes in statistical significance
of some real experiments. We show examples where
the estimated treatment effects change from statistically insignificant
to significant after incorporating prior information
in Table \ref{tab:insigTosig}, and vise versa in Table \ref{tab:sigToinsig}. In Table \ref{tab:insigTosig}, we see
that the treatment effect estimates are very close for many
experiment-treatment combinations, with the exception of 2
treatment groups ((A1, B1) and (A2, B1)). In this experiment,
different treatment groups correspond to different
combinations of A and B settings. Since the treatments are
similar, we expect that the impacts of these policies should
also be similar. By leveraging prior information we are able
to capture this similarity, whereas this similarity is ignored
without prior information.
Similarly, we see that in examples where the estimated
treatment effects become insignificant after incorporating
prior information (Table \ref{tab:sigToinsig}), the estimates based on
no prior information on M1 are much larger than the corresponding
estimates based on prior information. Specifically,
in this experiment, the average, minimum, and maximum
estimates based on no prior information on M1 in other
parameter settings (excluding treatment groups shown on
the table) are 16, -59, and 67, respectively, which are much
smaller than the estimates based on no prior information on
M1 for the treatment groups shown in Table \ref{tab:sigToinsig}.

\begin{table}[H]
\centering
  \caption{Estimated treatment effects on metric M4 that become
  significant after adopting using prior information. EST, CIL,
CLR correspond to the estimated
treatment effects, and their corresponding lower and
upper bounds of the 95\% credible intervals.}
  \begin{tabular}{c|ccc|ccc}
    \toprule
    \multirow{2}{*}{Treatment} &
      \multicolumn{3}{c|}{$k = \infty$} &
      \multicolumn{3}{c}{$k = 0$} \\
      & {EST} & {CIL} & {CIR} & {EST} & {CIL} & {CIR} \\
      \hline
    (A1, B2) & -12 & -45 & 21 & -14 & -25 & -4 \\
    (A1, B1) & 0.01 & -12 & 12 & -15 & -21 & -9 \\
    (A3, B1) & -11 & -26 & 3 & -16 & -22 & -10\\
    (A2, B1) & -0.14 & -14 & 14 & -17 & -24 & -11 \\
    \bottomrule
  \end{tabular}

\label{tab:insigTosig}
\end{table}
\vspace{-5mm}
\begin{table}[H]
\centering
  \caption{Estimated treatment effects on metric M1 that become
  significant after adopting using prior information. EST, CIL,
CLR correspond to the estimated
treatment effects, and their corresponding lower and
upper bounds of the 95\% credible intervals.}
  \begin{tabular}{c|ccc|ccc}
    \toprule
    \multirow{2}{*}{Treatment} &
      \multicolumn{3}{c|}{$k = \infty$} &
      \multicolumn{3}{c}{$k = 0$} \\
      & {EST} & {CIL} & {CIR} & {EST} & {CIL} & {CIR} \\
      \hline
    (A1, B3) & 145 & 8 & 282 & 3 & -22 & 29 \\
    (A2, B3) & 104 & 0.44 & 208 & 10 & -16 & 37 \\
    (A2, B2) & 113 & 31 & 195 & 13 & -9 & 37\\
    \bottomrule
  \end{tabular}

\label{tab:sigToinsig}
\end{table}

\section{Risks Under Different Trade-offs}
\vspace{5mm}
We illustrate how to leverage this framework for decision
making by answering the the following question from a policy
developer:

\begin{quote}
``I ran an experiment to assess the impact of my new
policy. Statistical results suggest that this policy
benefits some metrics but hurts some other metrics.
Should I launch the new policy? How would my
decision change based on my risk appetite?"
\end{quote}

To answer this question, we use two real experiments,
labeled E1 and E2, to illustrate the decision space. To simplify
the illustration, we ignore launch (roll-back) costs by
assuming $c_0 = c_1$. Therefore, $\mathcal{R}(a_1|\textbf{x}) < 0$
corresponds to
a launch decision and $\mathcal{R}(a_1|\textbf{x}) > 0$ to a
roll-back decision. In the decision space for the two experiments shown
in Figure \ref{fig:sameDay}-\ref{fig:rb} below, we consider two performance
metrics (viz., M1, M2) for E1 and two performance metrics (viz., M1, M3)
for E2 under different trade-offs. The decision space for E1 in Figure \ref{fig:sameDay}
suggests that regardless of the trade-offs, we should not
launch the policy since the risk is greater than 0 under all
reasonable trade-offs. This is intuitive since in general, we aim
to increase M1 while reducing M2, but the estimated impact
of this policy on M1 and M2 are -31 and 91, respectively.

\begin{figure}[H]
\centering
\includegraphics[width=8cm]{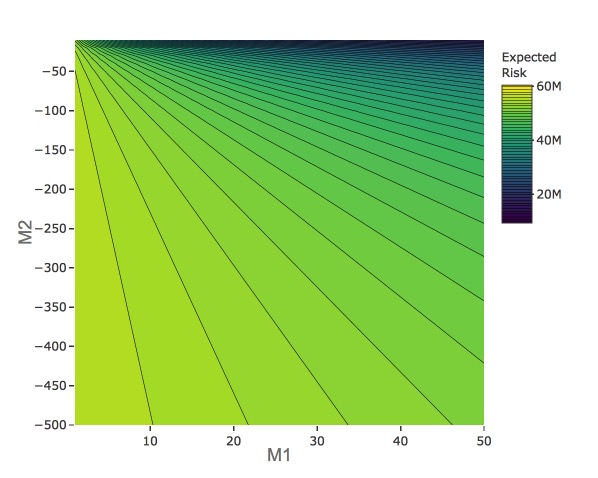}
\caption{Decision space for E1. $x$-axis and $y$-axis represent
the trade-offs between M1 and M2. For example, the
point corresponding to (20, -100) on the graph represent the
expected risk given trade-off vector (20, -100).}
\label{fig:sameDay}
\end{figure}

The decision space for E2 in Figure \ref{fig:rb} below suggests that
the launch decision depends on our
business preference (e.g., how much we are willing to invest
given an expected gain), which can be specified through a
trade-off vector. To make a decision in E2, we consider the
expected risk under a launch decision (e.g., if my policy
was rolled out, what is the expected risk?). Figure \ref{fig:rb} shows
the decision space (expected risk\footnote{The magnitude of the expected risk does not directly translate
to how much we save or lose, we focus on the sign of the
expected risk. We again assume that the operational cost of
maintaining the new policy is the same as that of the status
quo.}) under a roll-out decision.
The $y$-axis (M3) represents an unknown value of enhancing
customer experience. Since M3 is unknown, we choose a set
of values to see how the expected risk varies with this value.
A roll-out decision should be made when the expected risk is
low and below 0 (blue region in Figure \ref{fig:rb}) based on Amazon’s
business preference, and a roll-back decision should be made
otherwise. In Figure \ref{fig:rb}, we see that if we are willing to invest
100 or more in M1 for each unit gain in M3, then we should
launch the new policy. In addition, we see that the less we
are willing to invest in M1, the greater the value of M3 has
to be in order for rolling out this new policy to be beneficial
in the long run, which is intuitive.

\begin{figure}[H]
\centering
\includegraphics[width=8cm]{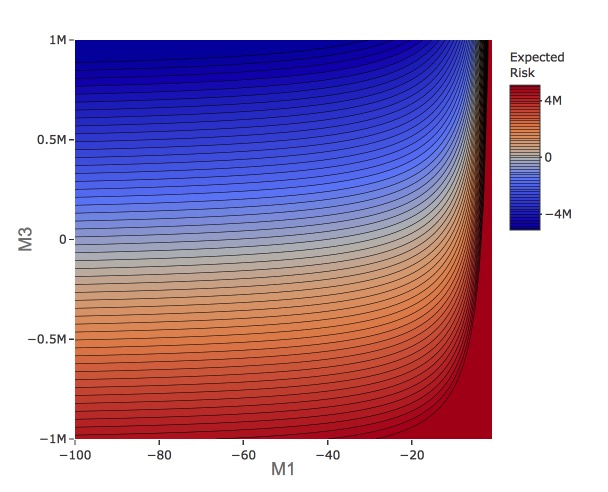}
\caption{A decision space (expected risk) under a roll-out
decision. The $y$-axis (M3) represents an unknown value
of enhancing customer experience. The $x$-axis represents
how much the business is willing to invest (e.g., a value of
-100 means that Amazon is willing to invest 100 in M1 for
each unit gain in M3). We vary the value of enhancing
customer experience ($y$-axis) and our risk appetite ($x$-axis)
to see how the expected risk (decision) varies. The blue
region corresponds to a roll-out recommendation (expected
risk < 0) while the red regions corresponds to a roll-back
recommendation.}
\label{fig:rb}
\end{figure}

The decision spaces presented above
provide decision makers intuitive illustration on how decisions
change with different loss functions. Given a loss function,
decision makers can quickly make launch decisions without
having to interpret any statistical results from experiments. 

\section{Defining Loss (Trade-offs) and automation}
\vspace{5mm}
In this section,
we discuss different approaches to defining a loss function
and highlight associated advantages and drawbacks. One way to define a
loss is to use business judgement. For example, a business
leader who owns the business believes that a \$1 increase in revenue is worth \$100
of investment in North America, the loss can simply be a trade-off vector representing
these beliefs. All new policies in North America should be subject to the same loss.
However, each business often involves multiple stakeholders who do not share the same
belief. Although this framework formalizes the fundamental objective of decision making, it
does not guarantee that stakeholders will agree on the same loss when presented with the
decision space. \\

To address this problem, another approach focuses on learning an `aggregated' business
preference through historical decisions. For example, suppose 1000 policies were assessed with 
experiments and associated decisions were recorded. As each launch decision corresponds
to a set of loss functions, the set of loss functions that lie in the the intersection of all the
launch sets will be the `aggregated' preference. However, this approach does not always produce a
loss function as the intersection can be empty. In addition, this approach leaves future
decisions susceptible to bad decisions made in the past. If only bad decisions were made in
the past, this approach does not allow decision makers to recover from bad decisions.\\

The final approach involves predicting the loss using economic indicators. Suppose our decision requires a trade-oﬀ
between revenue and customer experience (e.g., how often the
customer can get 2-day shipping when purchasing an item),
we can build a predictive model to estimate the revenue that
increasing fast-delivery availability will generate and use this
information to define a loss function. This approach oﬀers
an objective and data driven loss function. However, the
quality of the decision becomes dependent on the accuracy
of the predictive model which cannot be assessed in many
scenarios.

\section{Discussion}
\vspace{5mm}
We presented a framework for streamlined and scalable decision making under multiple objectives (performance metrics).
We showed that this framework can incorporate decision makers’ risk appetite and the costs of policies being assessed in
experiments in decision making. We also showed that leveraging learning from historical experiments increases treatment
eﬀect estimation eﬃciency while preserving coverage. Several
extension to this work can potentially reduce the costs of
experimentation and further increase the speed of Amazon’s
innovations. One limitation of our current framework is that
it requires the trade-oﬀs to be defined by stakeholders, which
can potentially introduce subjectively and ambiguity. Hence,
it is essential to identify a structured process to define these
trade-oﬀs. To make running experiments more cost eﬀective, we can further leverage historical learning and extend the
hierarchical structure to model treatment eﬀect evolution
during each experiment, and across experiments to determine
optimal stopping through sequential testing \cite{wald1973sequential}.

\section{Acknowledgements}
\vspace{5mm}
The authors thank David Afshartous, Luke Smith, Paavni
Rattan, Xiaolong Zhong, and Tarun Bhaskar for their insightful
feedback on this work.

\bibliographystyle{plain}
\bibliography{References.bib}

@book{gelman2013bayesian,
  title={Bayesian data analysis},
  author={Gelman, Andrew and Stern, Hal S and Carlin, John B and Dunson, David B and Vehtari, Aki and Rubin, Donald B},
  year={2013},
  publisher={Chapman and Hall/CRC}
}

@article{huggins2015risk,
  title={Risk and regret of hierarchical bayesian learners},
  author={Huggins, Jonathan H and Tenenbaum, Joshua B},
  journal={arXiv preprint arXiv:1505.04984},
  year={2015}
}

@book{berger2013statistical,
  title={Statistical decision theory and Bayesian analysis},
  author={Berger, James O},
  year={2013},
  publisher={Springer Science \& Business Media}
}

@article{blume2003your,
  title={What your statistician never told you about P-values},
  author={Blume, Jeffrey and Peipert, Jeffrey F},
  journal={The Journal of the American Association of Gynecologic Laparoscopists},
  volume={10},
  number={4},
  pages={439--444},
  year={2003},
  publisher={Elsevier}
}

@article{sterne2001sifting,
  title={Sifting the evidence?what's wrong with significance tests?},
  author={Sterne, Jonathan AC and Smith, George Davey},
  journal={Physical Therapy},
  volume={81},
  number={8},
  pages={1464--1469},
  year={2001},
  publisher={Oxford University Press}
}

@article{wasserstein2016asa,
  title={The ASA?s statement on p-values: context, process, and purpose},
  author={Wasserstein, Ronald L and Lazar, Nicole A and others},
  journal={The American Statistician},
  volume={70},
  number={2},
  pages={129--133},
  year={2016}
}

@book{gelman2006data,
  title={Data analysis using regression and multilevel/hierarchical models},
  author={Gelman, Andrew and Hill, Jennifer},
  year={2006},
  publisher={Cambridge university press}
}

@article{scott2015multi,
  title={Multi-armed bandit experiments in the online service economy},
  author={Scott, Steven L},
  journal={Applied Stochastic Models in Business and Industry},
  volume={31},
  number={1},
  pages={37--45},
  year={2015},
  publisher={Wiley Online Library}
}

@inproceedings{hill2017efficient,
  title={An efficient bandit algorithm for realtime multivariate optimization},
  author={Hill, Daniel N and Nassif, Houssam and Liu, Yi and Iyer, Anand and Vishwanathan, SVN},
  booktitle={Proceedings of the 23rd ACM SIGKDD International Conference on Knowledge Discovery and Data Mining},
  pages={1813--1821},
  year={2017},
  organization={ACM}
}

@book{wald1973sequential,
  title={Sequential analysis},
  author={Wald, Abraham},
  year={1973},
  publisher={Courier Corporation}
}

@article{thompson1933likelihood,
  title={On the likelihood that one unknown probability exceeds another in view of the evidence of two samples},
  author={Thompson, William R},
  journal={Biometrika},
  volume={25},
  number={3/4},
  pages={285--294},
  year={1933},
  publisher={JSTOR}
}

\end{document}